\documentclass[aps,prl,twocolumn,showpacs,superscriptaddress,groupedaddress]{revtex4}  
\usepackage{graphicx}  
\usepackage{dcolumn}   
\usepackage{bm}        
\usepackage{amssymb, amsmath, amsfonts}   
\usepackage{lipsum}        

\DeclareGraphicsExtensions{{.pdf}}
\graphicspath{{./}}

\begin{document}

\widetext


\title{Purely hydrodynamic origin for swarming of swimming particles}
\author{Norihiro Oyama}
\author{John Jairo Molina}
\author{Ryoichi Yamamoto}
\affiliation{Department of Chemical Engineering, Kyoto University, Kyoto 615-8510, Japan}
\date{\today}

\begin{abstract} Three-dimensional simulations with fully resolved
 hydrodynamics are performed to study the collective motion of model
 swimmers in confinement. We show that certain swimming
 mechanisms can lead to traveling wave-like collective motion even
 without any direct alignment mechanism. It is also shown that by
 varying the swimming mechanism, this collective motion can be
 suppressed, contrary to the perception that hydrodynamic effects are
 completely screened at high volume fraction. From an analysis of bulk systems, it is shown
 that this traveling wave-like motion, which can be characterized as
 a pseudo-acoustic mode, is mainly due to the intrinsic swimming
 property of the particles.
 
\end{abstract}

\maketitle


Active matter systems such as groups of microorganisms, birds or fish often form swarms or flocks,
and typically show complex collective behavior~\cite{Sokolov:2007td,Nash:2010ir,Mishra:2014hy}.
Of special interest for scientific study are active microscopic systems,
e.g. microorganisms or Janus particles, because they are
particularly well suited for experiments.
Understanding the collective dynamic behavior of
these systems can be useful
for both science and industrial applications,
for example, to explain the formation of biofilms~\cite{Nadell:2009de}
or to design targeted drug delivery systems~\cite{Laage:2006jw}.
Such systems typically exist under some type of confinement, which has
been shown to exert a strong influence on the collective
behavior~\cite{Angelani:2009kk,DiLeonardo:2010kk,Woodhouse:2012cl,Kaiser:2012ur,Kaiser:2014bn}.
So far, most analysis of such systems have either neglected hydrodynamic interactions or
only considered them in a far field approximation~\cite{AditiSimha:2002eg,Angelani:2009kk,Kaiser:2012ur,Ezhilan:2013ec}, even though it
is known that the hydrodynamic interactions can dramatically affect the
physical properties of the systems~\cite{Rafai:2010cn}.
Theoretically, it's impossible to take into account the full
hydrodynamic interactions in many body systems, and numerically it's very computationally expensive.
While several studies have succeeded in including hydrodynamic
interactions~\cite{Evans:2011cw,Alarcon:2013dg,Li:2014we},
none have focused on the collective dynamic behavior of 3D active
systems under confinement.
The recent work by Z\"ottl~and~Stark\cite{Zottl:2014wy} is related to ours, but deals with
pseudo-2D systems under strong confinement (small separation of
walls), high volume fraction and strong swimming strength, which are
quite different conditions from those of the present work.
Not surprisingly, we are able to observe quite different dynamic behavior.

We have conducted direct numerical calculations for 3D systems of
spherical swimmers in a host viscous fluid
with fully resolved hydrodynamics.
Under the confinement of two flat parallel walls,
we find collective motion of swimmers with traveling wave
characteristics, which propagates back and forth between the walls.
Although similar traveling wave-like motion has already been observed in
systems with explicit alignment interactions between the particles~\cite{Chate:2008ca,Ohta:2014dp},
here we show that hydrodynamic interactions alone are enough to
exhibit this behavior.
We also confirmed that this traveling wave-like behavior is not due to
the presence of the walls,
but is just a manifestation of the bulk behavior.
Crucial to observe this behavior are the strength and type
(pushers vs. pullers)  of swimming, not the degree of
global alignment.

As the numerical model of microswimmers,
the Squirmer Model was employed.
In this model, a self-propelled particle is modeled
as a spherical object with a modified stick boundary condition at its surface~\cite{Lighthill:1952ta,Blake:2006ig}:
\begin{eqnarray}
\boldsymbol{u}^{s} \left( \hat{\boldsymbol{r}} \right) = \sum_{i=1}^\infty\frac{2}{n(n+1)}B_n P_n^\prime(\cos{\theta})
\sin{\theta}\hat{\boldsymbol{\theta}}
\label{Sq_full},
\end{eqnarray}
where $\hat{\boldsymbol{r}}$ is a unit vector directed from the
center of a squirmer to a point on its surface,
$\theta=\cos^{-1}{(\hat{\boldsymbol{r}}\cdot\hat{\boldsymbol{e}})}$
the polar angle between $\hat{\boldsymbol{r}}$ and the swimming direction $\hat{\boldsymbol{e}}$,
 $\hat{\boldsymbol{\theta}}$ the unit polar angle vector.
$P_n^\prime$ is the derivative of the Legendre polynomial of $n$-th
order, and $B_n$ is the magnitude of each mode.
Here, the radial deformation has been ignored, 
so this surface velocity has only the tangential component, and is
responsible for the self-propulsion of the swimmers.

\begin{figure}
\includegraphics[width=\linewidth]{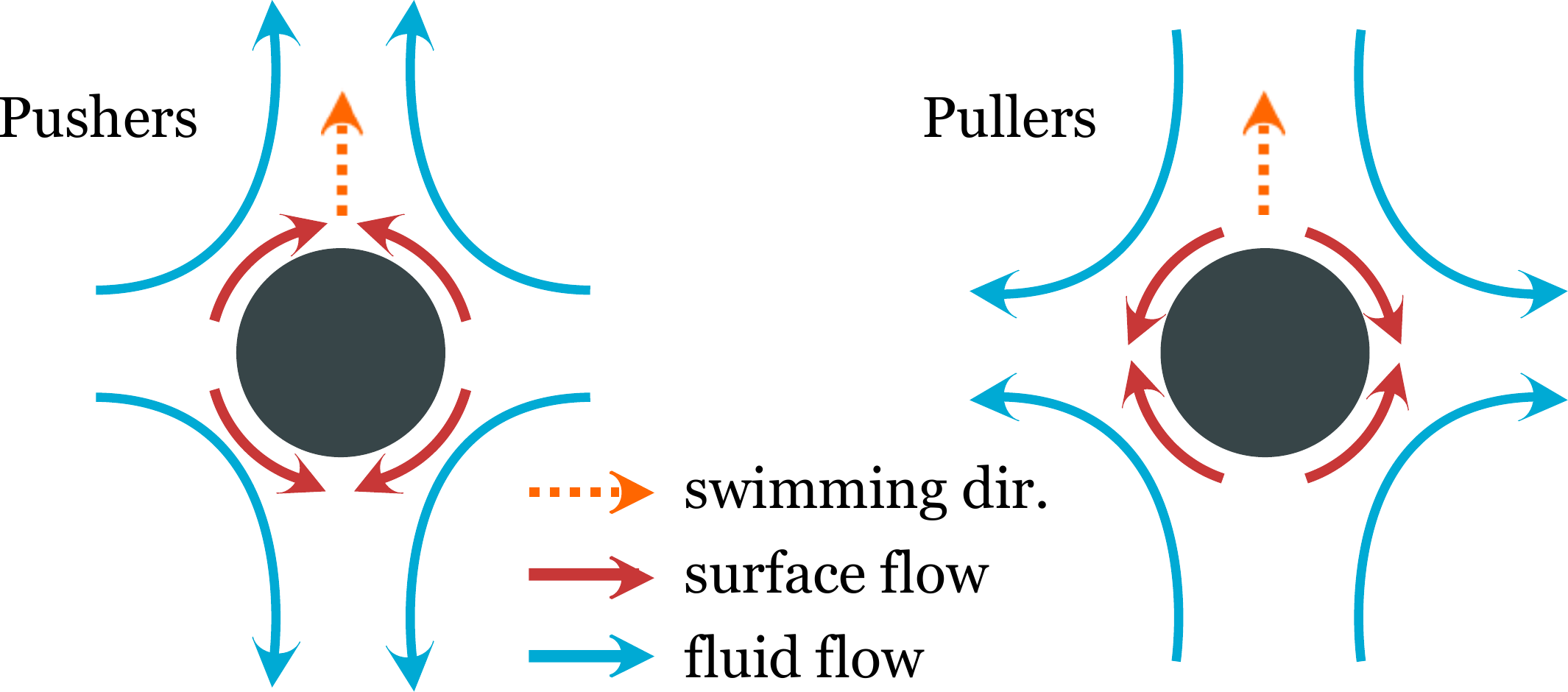}
\caption{\label{fig:squirmer}  Schematic representation of pushers and
  pullers within the squirmer model. The swimming direction, and
  the fluid and surface flows generated by squirmers
  are shown.}
\end{figure}

In this work, only the first two modes in Eq.(\ref{Sq_full}) were retained:
\begin{eqnarray}
\boldsymbol{u}^s(\theta) = B_1\left(\sin{\theta} + \frac{\alpha}{2}\sin{2\theta}\right)\hat{\boldsymbol{\theta}}
\label{Sq_2}.
\end{eqnarray}
The coefficient of the first mode, $B_1$, determines the swimming velocity of an isolated squirmer, as $U_0=2/3B_1$.
The coefficient of the second mode, $B_2$, determines the strength of stirring, or the stresslet~\cite{Ishikawa:2007cf}.
Using only the first two modes, the type of swimming can be modified by changing the sign of the second coefficient, 
$\alpha$ in Eq.(\ref{Sq_2}).
Negative values of $\alpha$ describe pushers, which generate an
extensile flow field in the propeling direction, and positive values describe
pullers, which generate a contractile flow (Fig.~\ref{fig:squirmer}).
Because these two swimming mechanisms result in distinct
flow fields, different swimmers can exhibit vastly different
collective motion, as we will show in this paper.

In this study, the squirmer model is incorporated into the Smoothed Profile Method (SPM)~\cite{Nakayama2005a,Kim:2006io,Nakayama:2008fi}, which is 
an efficient calculation scheme for solid/fluid two phase dynamics problems with full hydrodynamics.
In this method, the sharp boundary between the solid and fluid domains
is not considered explicitly.
Instead, a diffuse interface of finite width $\zeta$ is introduced,
and the solid domain is represented via an order parameter $\phi$,
which takes a value of 1 in the solid domain, and 0 in the
fluid domain.
Using this continuous order parameter, all the physical quantities can
be expressed as field variables on Cartesian grids, and the calculation cost can be reduced drastically.
As the governing equations for the total fluid, we employ a modified
incompressible Navier-Stokes
equation:
\begin{align}
    \rho_{\text{f}} \left( \partial_t + \boldsymbol{u}\cdot\nabla\right)\boldsymbol{u}
                &= \nabla \cdot \boldsymbol{\sigma} + \rho_{\text{f}}\left( \phi\boldsymbol{f}_{\text{p}} +\boldsymbol{f}_{\text{sq}} \right)\notag,\\
     \boldsymbol{\nabla}\cdot\boldsymbol{u} &= 0,
       \\\boldsymbol{u}&=(1-\phi)\boldsymbol{u}_{\text{f}}+\phi
                \boldsymbol{u}_{\text{p}}\notag,\\
                \phi \boldsymbol{u}_{\text{p}} &= \sum_{i}\phi_i
                     [\boldsymbol{V}_i+\boldsymbol{\Omega}_i \times \boldsymbol{R}_i]\notag,
   \end{align}
where $\boldsymbol{u}$ is the total velocity field,
$(1-\phi)\boldsymbol{u}_{\text{f}}$ and $\phi\boldsymbol{u}_{\text{p}}$ the
velocity field of fluid and particles, respectively.
The position, velocity and  angular
velocity of particle $i$ are expressed as $\boldsymbol{R}_i$,
$\boldsymbol{V}_i$ and $\boldsymbol{\Omega}_i$.
The stress tensor and the mass density of the host fluid are
$\boldsymbol{\sigma}$, $\rho_\text{f}$.
The body force generated from the
rigidity condition of particles is calculated as $\phi\boldsymbol{f}_{\text{p}}$, and
$\boldsymbol{f}_{\text{sq}}$ stands for the force field resulting from the squirming motion~\cite{Molina:2013hq}.
The time evolution of the particles follows from the Newton-Euler equations:
  \begin{align}
                \dot{\boldsymbol{R}}_i &= \boldsymbol{V}_i ,
                &\dot{\boldsymbol{Q}}_i &= \text{skew}(\boldsymbol{\Omega}_i)\cdot
                \boldsymbol{Q}_i,\\
                M_{\text{p}}\dot{\boldsymbol{V}}_i &= \boldsymbol{F}_i^{\text{H}}+\boldsymbol{F}_i^{\text{C}},
                &\boldsymbol{I}_{\text{p}}\cdot\dot{\boldsymbol{\Omega}}_i &= \boldsymbol{N}_i^{\text{H}}\notag
               , \end{align}
  where $\boldsymbol{Q}_i$ denotes the
  rotation matrix,  $M_{\text{p}}$ the mass,
$\text{skew}(\boldsymbol{\Omega}_i)$ the skew symmetric matrix of
  $\boldsymbol{\Omega}_i$, and  $\boldsymbol{F}_i^{\text{H}}$ and
$\boldsymbol{N}_i^{\text{H}}$ are the hydrodynamic force and torque acting
  on particle $i$.
The force due to direct interactions between particles, $
\boldsymbol{F}_i^{\text{C}} $, is taken to be a truncated Lennard-Jones potential, 
in order to ensure that the excluded volume constraint is satisfied.
See refs~\cite{Nakayama2005a,Kim:2006io,Nakayama:2008fi,Molina:2013hq}
for more detailed information of this method.

Using this computational model, 
we conducted simulations of squirmer dispersion systems in a three dimensional domain.
Throughout this paper, the diameter $a$ and the boundary thickness
$\zeta$ of the squirmers are $4\Delta$, $2\Delta$,  where $\Delta$ is the grid spacing and the unit length, and the parameter $B_1$ in Eq.~(\ref{Sq_2}) is 0.375, respectively.
The shear viscosity $\eta$ and the fluid mass density
$\rho_{\text{f}}$ are set to one, 
and the unit time is then expressed as $\rho\Delta^2 /\mu$.
The particle Reynolds number of an isolated squirmer $Re^0 = \rho_{\text{f}}U_0a/\eta$
is set equal to one. The Reynolds number in the dispersion will be
different from $Re^0$ due to a decrease in the propelling velocity with
increasing volume fraction.
We have ignored any effects due to thermal fluctuations.
No external force is exerted, and the systems are buoyancy free.

\begin{figure}
   \includegraphics{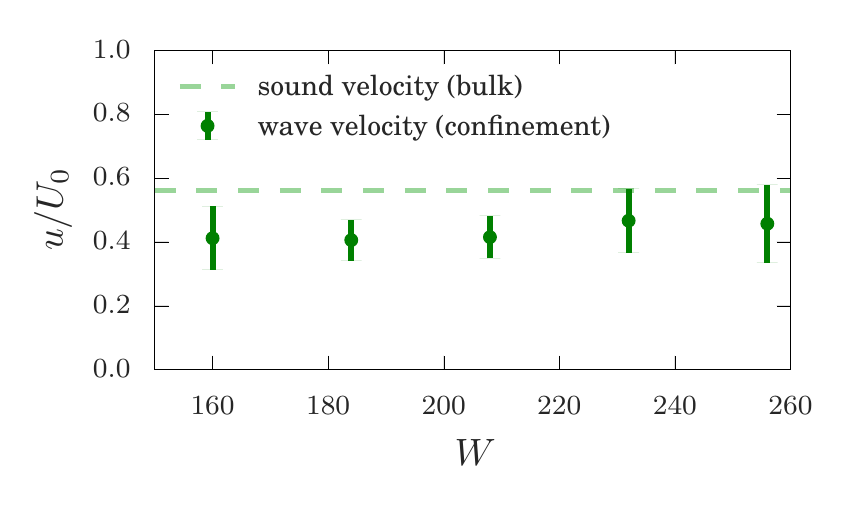}
   \caption{\label{fig:vel_wave}
     The traveling wave velocity of confined systems as a function of wall separation $W$.
     The sound velocity of the bulk system is also shown. Here, only
     puller results are shown, as pushers don't show wave-like behavior.
     All the values are normalized by $U_0$.
   }
\end{figure}

We first calculated the dynamics of $N=2607$ swimmers in a $64\Delta\times160\Delta\times64\Delta$ rectangular system
(volume fraction of 13\%).
 The computational domain is periodic in the $x$ and $z$ direction,
 but confined between two flat parallel walls along the $y$ direction; at $y=1\Delta$ and $159\Delta$.
 These walls are made of $2048$ spherical particles which are pinned and unable to move or rotate from the initial positions and orientations.
The wall particles are regularly spaced and are defined to have the
same diameter as the squirmers.
The initial configuration of swimmers was randomly determined with
respect to both position and orientation.
We conducted simulations for various $\alpha$ values in order to
investigate the effect of $\alpha$ on the dynamics.
We have calculated  the time evolution of the local
density $\rho$, polar order $Q_\text{l}$ and velocity $v_\text{l}$, as a function of distance from the
walls.
Here, $Q_\text{l}=\langle \hat{e}_{y}\rangle_{\text{plane}}$ and $\bar
{v}_\text{l}= \langle {V}_{y}/U_0\rangle_{\text{plane}}$, where $
\hat{e}_{y}$ and $V_y$ are the components of $\hat{\boldsymbol{e}}$ and $\boldsymbol{V}$ perpendicular
to the walls, and $\langle \cdot\rangle_{\text{plane}}$ denotes an
ensemble average over all particles in parallel slabs of width $2\Delta$.
Fig.~\ref{fig:density_fluc} shows the results of the systems with
$\alpha=\pm 0.5$ (see supplementary materials S1 and S2).
For pullers ($\alpha =0.5$), the time evolution of the density, polar
order and velocity are shown,
while for pushers ($\alpha =
-0.5$), only density is shown.
Only the pullers show large density fluctuations, which we note are due to hydrodynamic
interactions alone, as no direct alignment interaction has been
included; pushers show no such behavior.
First, the pullers build up two distinct traveling waves
which bounce back and forth between the walls.
Gradually, as one wave begins to dominate, they merge into
a single propagating wave.
We also considered different wall separations, and found that such
propagating wave-like motion can only be observed if the wall
separation is larger than the characteristic size of this propagating
flock.
In our simulations, the threshold is $W/a\simeq40$, where $W$ is the
wall separation.
In fig.~\ref{fig:vel_wave}, the traveling wave velocity is plotted for
several different wall separations (the bulk sound velocity which will be explained later is also shown).
They are quantitatively the same and there is no clear separation dependence.
From these measurements,
this traveling wave-like motion can be understood as an intrinsic property of the
self-propelled systems, which is independent of the wall effects.

\begin{figure}
   \includegraphics{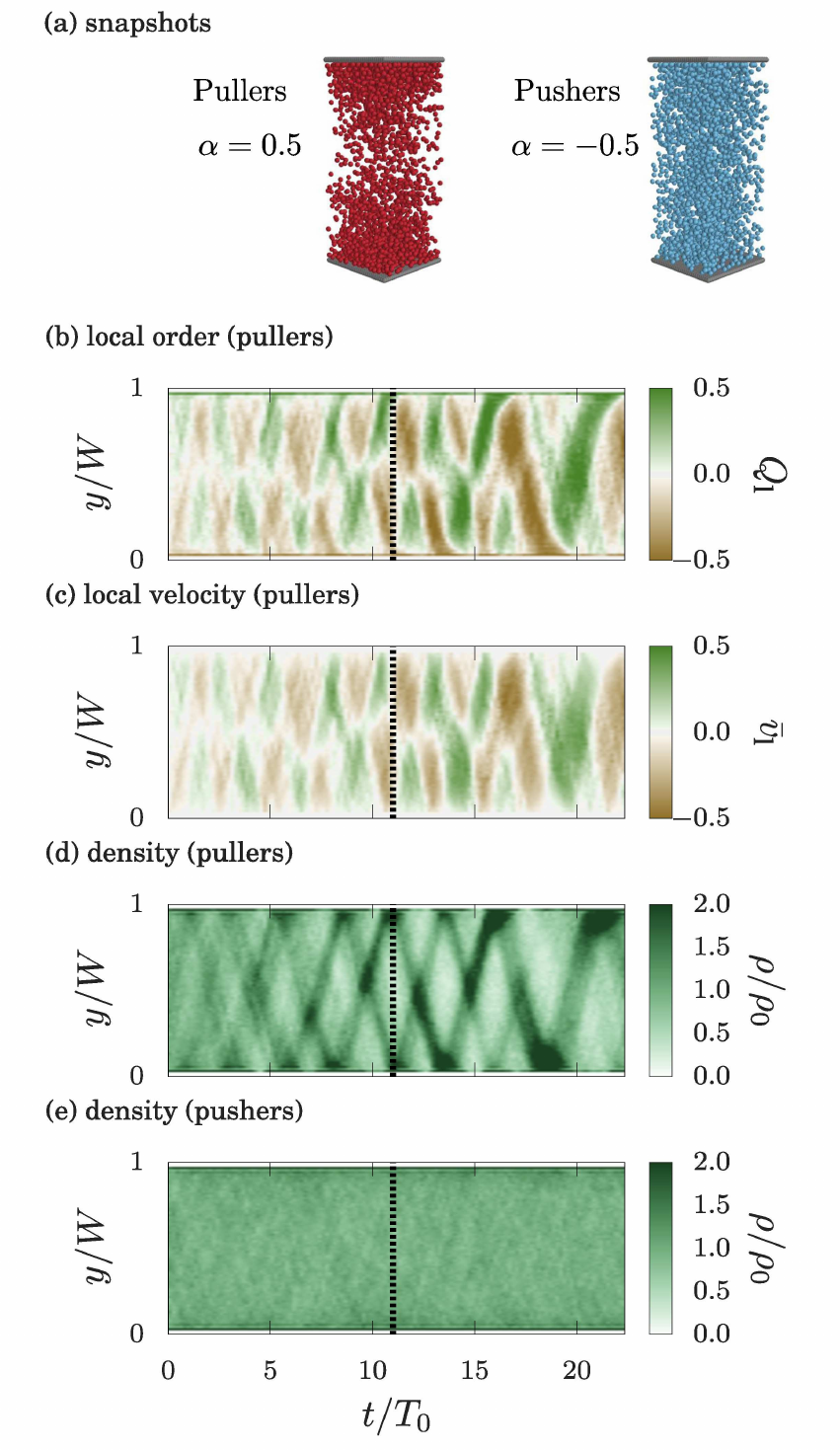}
   \caption{\label{fig:density_fluc} Simulation results for pullers
     and pushers ($\alpha=\pm 0.5$) at volume fraction of 13\% and
     wall separation $W/a=40$.
     (a) Simulation snapshots at $T=11T_0$,
     (b) local order of pullers defined as  $Q_\text{l}=\langle
     \hat{e}_{y}\rangle_{\text{plane}}$,
     (c) local velocity of pullers in y direction (perpendicular to the walls)
     defined as $\bar {v}_\text{l}= \langle
     {V}_{y}/U_0\rangle_{\text{plane}}$
     (d,e) plane averaged density for pushers and pullers at each
     height $y/W$.
    The density is normalized by the global
    average value, $\rho_0$, and time and velocity are normalized by
    $T_0=W/U_0$ and $U_0$, where $W$
    represents the separation of parallel flat walls.
   The snapshots are of around $t=11T_0$, which is marked by dashed
   line.
   }
\end{figure}

To confirm this conjecture,
further simulations were conducted in cubic systems with no
confinement, under full periodic boundary conditions in all directions.
The linear dimension of the computational domain is $L=64\Delta$ and
the number of swimmers is $1043$ ( volume fraction of 13 \%, the same value as the previous simulations in the
systems with walls).
In order to study the density correlations, we calculated the dynamic
structure factor $S \left( \boldsymbol{k} , \omega \right)$, which is just the Fourier transform (in both space and time)
of the density-density correlation function\cite{Hansen}:
\begin{align}
\rho_{ \boldsymbol k}(t)&=\sum_{i=1}^{N}\exp\left[ -i\boldsymbol{k}\cdot\boldsymbol{R}_i(t)\right]
\\
S \left( \boldsymbol k, \omega\right) &=
\frac{1}{2\pi}\int_{-\infty}^{\infty} \frac{1}{N}\langle \rho_{\boldsymbol k} \left(
t\right)\rho_{- \boldsymbol k} \left( 0 \right)\rangle  \text{exp} \left( i\omega t \right)
\text{d}t
\end{align}
where $\boldsymbol{k}$ is the
wave vector and $\omega$ the angular frequency.

\begin{figure}
\includegraphics{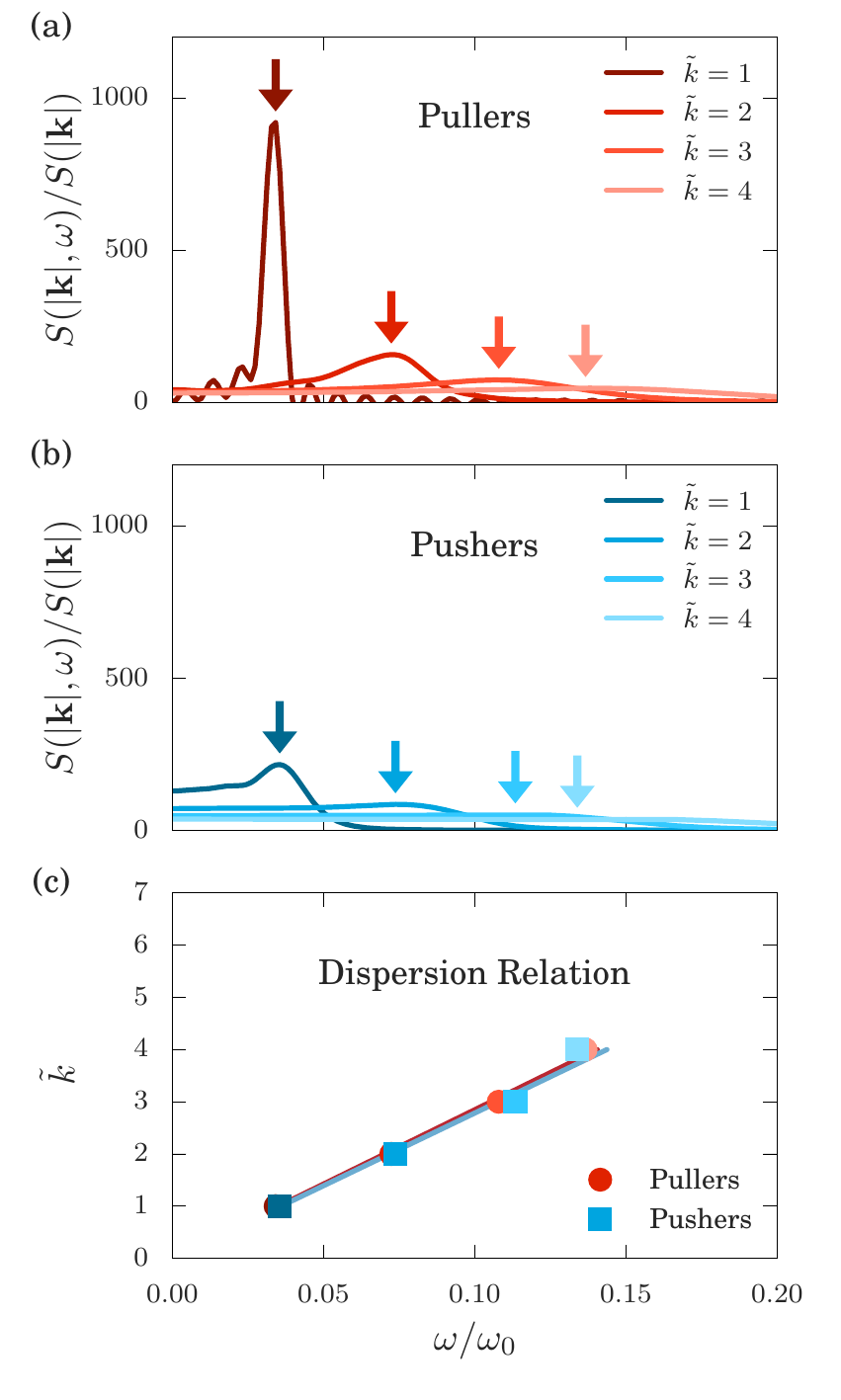}
\caption{\label{fig:DSF}  The calculation results for the spherically
  averaged Dynamic Structure
  Factor (DSF) $S \left( \boldsymbol k, \omega\right)$,
  where $\boldsymbol k =\frac{2\pi}{L}\tilde{ \boldsymbol{k}}$ is the
  wave vector, and $\omega$ the angular frequency (normalized by $\omega_0=2\pi/t_0$, where $t_0 = a/U_0$). 
The top two figures show the simulation results for the DSF of (a)
pullers at $\alpha =
0.5$, and (b) pushers at $\alpha = -0.5$.
The bottom figure (c) shows the relation between wave vectors and the
positions of the Brillouin peaks.
Arrows in (a) and (b) show the positions of peaks.}
\end{figure}

The spherically averaged (over shells of wave vectors) structure factors
are shown in Fig.~\ref{fig:DSF}.
Shifted peaks  were observed for both systems, pullers and pushers.
These peak positions $\omega_{\text{shift}}$ follow a Brillouin-like dispersion relation:
\begin{align}
  \omega_{\text{shift}} = c|\boldsymbol{k}|
  \label{eq:Dis}
\end{align}
where $c$ is a coefficient which has the dimension of velocity (in
the case of normal Brillouin peaks, $c$ represents the sound
velocity).
We can therefore say that these systems possess pseudo-acoustic mode dynamics,
or a sound-like propagation mode for the density.
No Rayleigh-like peaks were observed for small wave
vectors, probably because the Brillouin-like peaks are much stronger
(for wave vectors of large magnitude, Rayleigh-like peaks are recovered).
The sound velocity of this pseudo-acoustic mode can be calculated from
the dispersion relation,shown at the bottom of
Fig.~\ref{fig:DSF}.
We don't see any difference between the sound velocities of pushers
and pullers, although the peak intensities show very big differences.
The calculated value of the sound velocity for the pullers is quantitatively the same as the 
propagating wave velocities in the confined systems, as shown in
fig.~\ref{fig:vel_wave}, although the velocities are slightly reduced
for small wall separations (Supplementary Material S3).
Therefore, we can conclude that the traveling wave-like behavior in
the confined system
is mainly due to the 
intrinsic pseudo-acoustic property of the
swimmers themselves, and the wall effects are not crucial.

Now we would like to discuss the origin of the pseudo-acoustic mode.
From fig.~\ref{fig:density_fluc}, we can see that the particles show local polar order and
move as a group collectively, even though individual particles move
independently of each other (for individual particles motion,
see Supplemental Material S4). In puller systems, this cooperative
swimming as a group can be understood as the origin of the
pseudo-acoustic mode, or density propagation.
Now let us consider
a simplified set-up
in which swimmers are placed along a straight line with a non-regular distribution,
and oriented parallel to this line, in such a way they will start
swimming in the same direction (Supplemental Material S5 and S6).
In this situation, due to the contractile flow along the swimming
direction, pullers attract each other and tend to enhance the
longitudinal mode of density fluctuations, resulting in the formation
of flocks. On the other hand, pullers show the opposite tendency, which
suppresses the longitudinal fluctuations.
In the systems of interest in this paper, the situation is much
more complicated.
However, we can still say that 
the appearance of the large density fluctuations we
present depends on the local interaction rather than on the global
order.
Actually, we never observed large fluctuations for pushers, even
though the global polar order
($Q=\langle \frac{1}{N}|\sum_{i} \hat{\boldsymbol{e}_i}|
\rangle$)~\cite{Evans:2011cw,Alarcon:2013dg}
could be similar to pullers with 
$\alpha=0.5$.
On the other hand, pullers with
various values of $\alpha$
($0.3\leq\alpha\leq0.9$) do show traveling-wave like motion in confinement.
In order to rule out the possibility that this fluctuation is due to
inertial effects, we also conducted a simulation with smaller value of
Reynolds number ($Re_0=0.1$). 
The system showed the same collective behavior. We can therefore
conclude that this
behavior is purely
hydrodynamic in origin.

Several articles have shown similar results; some are consistent with
our findings, while others are not.
Simha and Ramaswamy predicted, using a continuum model that active
matter suspensions with polar order can show sound like
propagation modes\cite{AditiSimha:2002eg}.
They mainly studied the propagation mode of a director field, but also
showed that the concentration field can exhibit the same type of propagation.
Ezhilan {\it et al.} studied a similar continuum model~\cite{Ezhilan:2013ec}.
They reported pushers show stronger instability in their concentration field,
contrary to our results that pullers have much higher Brillouin-like peaks and are more likely to show density fluctuations.
They employed a far-field approximation, therefore they did not consider the
finite size of the particles or the full hydrodynamic interactions.
In our study, we took into account the full hydrodynamic interactions and finite
volume of the swimmers.
This difference is likely the cause for the contradictory results.
Alarcon and Pagonabarraga\cite{Alarcon:2013dg} reported the same kind
of density fluctuations in bulk puller systems, using direct numerical
simulations with the same model swimmer system.
They focused on the flocking behavior in bulk systems, while we studied the behaviors in confinement, which
makes the flocking more apparent, and also identified the pseudo-sound
property of the systems. 

In conclusion, we confirmed, using direct numerical simulations
with full hydrodynamics, that swimmer dispersions confined between flat parallel walls
can exhibit a traveling wave-like motion. Furthermore, this non-trivial
collective behavior is mainly due to the pseudo-acoustic
properties of the swimmers themselves, with a purely hydrodynamic origin.
This kind of phenomenon cannot be expected in conventional passive
colloidal dispersions, since such density fluctuations would
be suppressed due to viscous damping\cite{Hess:2006bo}.
In future works, we will consider complex wall geometries, such as
gear shaped objects~\cite{Kaiser:2012ur,Angelani:2009kk}
and circular systems~\cite{Woodhouse:2012cl}.
Based on our current results, we expect that spherically shaped particles, without any
apparent alignment interaction, can generate non trivial collective
motion due solely to hydrodynamic interactions.

We thank Simon K. Schnyder for useful discussions.
This work was supported by JSPS KAKENHI Grant Number 26247069 and also
by a Grant-in-Aid for Scientific Research on Innovative Areas
``Fluctuation \& Structure" (No.26103522) from the Ministry of
Education, Culture, Sports, Science, and Technology of Japan.
We also acknowledge the supporting program for interaction-based initiative team studies (SPIRITS) of Kyoto University.



\end{document}